\documentclass{mn2e}
\usepackage{psfig}
\usepackage{epsfig}

\newif\ifAMStwofonts

\def\sqiglt{\hbox{\rlap{\lower.55ex \hbox {$\sim$}}\kern-.05em \raise.4ex \hbox{$<$}\,}}
\def\sqiggt{\hbox{\rlap{\lower.55ex \hbox {$\sim$}}\kern-.05em \raise.4ex \hbox{$>$}\,}}

\def\til{\ensuremath{\sim\,}}
\def\w{\ensuremath{\omega}}
\def\p{\ensuremath{\phi}}
\def\W{\ensuremath{\Omega}}

\def\beat{\mbox{\w\,--\,\W}}
\def\chisq{\ensuremath{\chi^2}}
\def\rchisq{\ensuremath{\chi_{\nu}^{2}}}

\newcommand{\tim}[1]{\ensuremath{\times 10^{#1}}}
\def\deg{\ensuremath{^{\circ}}}
\def\etal{et al.\ }

\title[FO~Aquarii]{Twisted accretion curtains in the intermediate polar 
FO Aquarii}
\author[P.\,A. Evans et al.]{P.\,A. Evans$^1$, Coel Hellier$^1$,
Gavin Ramsay$^{2}$ \&\ Mark Cropper$^{2}$\\
$^1$Astrophysics Group, School of Chemistry and Physics, Keele University, 
Staffordshire, ST5 5BG\\
$^2$Mullard Space Science Laboratory, University College London, Holmbury 
St.~Mary, Dorking, Surrey RH5 6NT}

\date{Accepted 
      Received }

\pagerange{\pageref{firstpage}--\pageref{lastpage}}
\pubyear{2003}

\begin{document}

\maketitle

\label{firstpage}

\begin{abstract}
We report on a \til37-ks \emph{XMM-Newton\/} observation of the intermediate
polar FO Aquarii, presenting X-ray and UV data from the EPIC and OM cameras. 
We find that the system has changed from its previously reported state of
disc-overflow accretion to one of purely disc-fed accretion.  We detect the
previously reported `notch' feature in the X-ray spin pulse, and explain it as
a partial occultation of the upper accretion pole.  Maximum flux of the
quasi-sinusoidal UV pulse coincides with the notch, in keeping with this idea. 
However, an absorption dip owing to the outer accretion curtains occurs 0.27
later than the expected phase, which implies that the accretion curtains are
twisted, trailing the magnetic poles. This result is the opposite of that
reported in PQ~Gem, where accreting field lines were found to lead the pole. 
We discuss how such twists relate to the accretion torques and thus the
observed period changes of the white dwarfs, but find no simple connection.
\end{abstract}


\section{Introduction}
\label{sec:intro}

\begin{figure*}
\begin{center}
\psfig{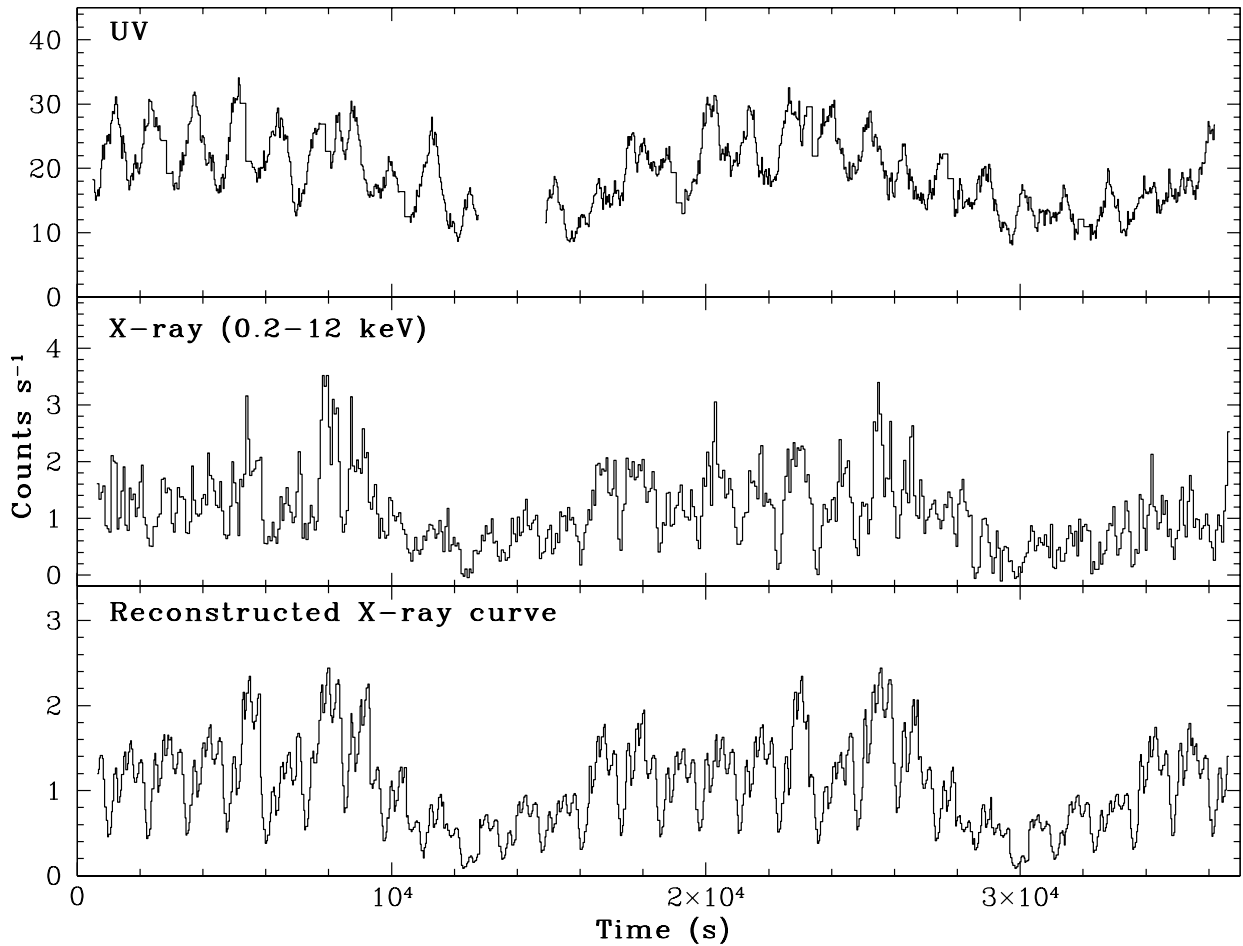}
\caption{The UV lightcurve (taken with the OM observing through the
UVM2 filter and binned at 32 s; top panel) and the EPIC-MOS X-ray
lightcurve (binned at 64 s; centre panel).  The bottom panel contains
the reconstructed model of the X-ray lightcurve, consisting of only
spin-cycle and orbital modulations (see text).}
\label{fig:curve}
\end{center}
\end{figure*}

FO~Aqr is an intermediate polar (IP), a magnetic variant of the
cataclysmic-variable class of binary stars. These systems consist of
a Roche-lobe-filling, late-type secondary and a white-dwarf primary,
the latter having a strong magnetic field (\til1--10 MG), inclined to
the rotational axis of the star. The in-falling matter forms an
accretion disc which is truncated at the magnetospheric boundary,
where it threads onto the white dwarf's magnetic field. The material
is then channeled into `accretion curtains' which guide the material
towards the poles, at which stand-off shocks form, heating material
to \til10$^8$ K (see Patterson 1994 for an in-depth review).

The accretion geometry of FO~Aqr has been the subject of much debate
(e.g.\ Hellier 1991; Norton \etal1992; Hellier 1993; Beardmore
\etal1998; the last three hereafter referred to as N92, H93 and B98
respectively),  but it is now generally accepted that the system
shows disc-overflow accretion, with most of the material accreting
via disc-fed accretion curtains (causing a pulse at the 20.9-min spin
cycle), and a component of the accretion stream overflowing the
disc and coupling to the magnetic field directly (causing a pulsation
at the beat period between the spin and 4.85-hr orbital cycles).

Whereas many IPs have quasi-sinusoidal spin pulses, FO~Aqr shows a
more complex pulse, including a broader dip and a narrow `notch'
feature, both of which are variable over time (N92, B98). Our aim
here is to use the superior spectroscopic capability of the
\emph{XMM-Newton\/} satellite to investigate the spin pulse and thus
the way in which material leaves the accretion disc and threads onto
field lines.

\subsection{The spin-cycle ephemeris}

The white dwarf in FO~Aqr was spinning down during the
1980s (Shafter \&\ Macry 1987), but Steiman-Cameron, Imamura \&\
Steiman-Cameron (1989) reported that the period-lengthening had
almost stopped by the time of their 1987 observations. Osborne \&
Mukai (1989) and Patterson \etal (1998) found that FO~Aqr then
began spinning up, which is confirmed by the most recent ephemeris by
Williams (2003).  However, these ephemerides suffer from significant
$O-C$\/ jitter and develop a cycle-count ambiguity for times later than
1998 (Williams 2003).  Thus we cannot compare the phasing of our data
(from 2001) with previous reports with any confidence.

\section{Observations}
\label{sec:obs}

FO~Aqr was observed by the \emph{XMM-Newton\/} satellite (Jansen et
al. 2001) for 37 ks on 2001 May 12. The X-ray MOS-1, MOS-2 (Turner
\etal2001) and PN (Str\"{u}der \etal2001) cameras were operating in
{\sc small window mode}, using the {\sc medium} filter. The optical/UV OM
camera (Mason et al. 2001) was operating in {\sc fast mode},
observing through the UVM2 filter (2050--2450 \AA). The RGS cameras
were also in operation; however the count-rate was too low to allow
phase-resolved analysis, which is the goal of this paper, thus they
are only mentioned in passing here. Some aspects of these data were
previously reported by Cropper \etal(2002).

We analysed the data using the {\sc xmm-sas} software v5.4.1. The
source data were extracted from a circular region of radius 3.5 arcsec
enclosing the source, with the background being taken from an annulus
around this area. Only single or double pixel events with a zero
quality flag were selected.

\section{Lightcurves and Power Spectra}
\label{sec:fts}

The X-ray and UV lightcurves are presented in Fig.~\ref{fig:curve},
binned at 64 s and 32 s respectively. We show the sum of the MOS
lightcurves; the PN lightcurve is very similar to the MOS curves and
so is not shown. In any case, the flickering level is higher than the
photon noise.

\begin{figure}
\begin{center}
\psfig{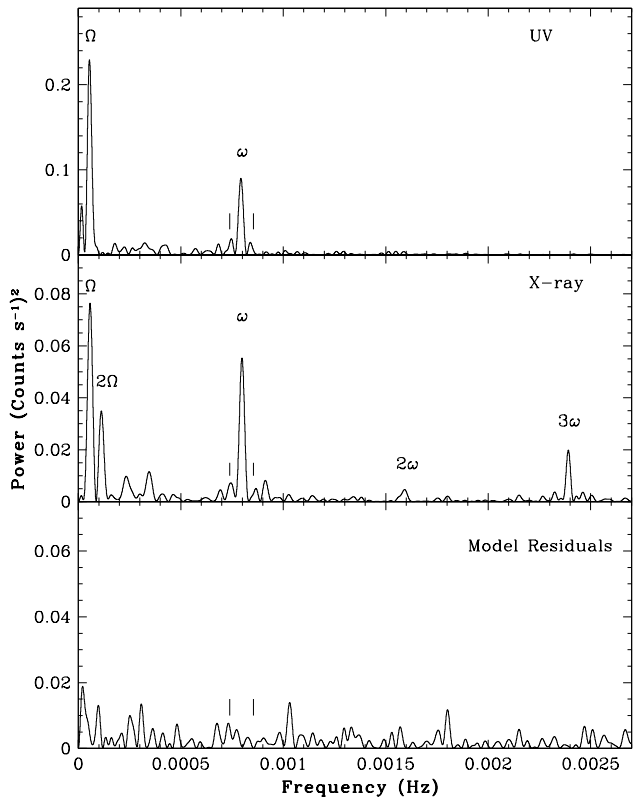}
\caption{Power spectra of the UV and X-ray (MOS 1+2) data and the model
residuals. \W\ and \w\ are the orbital and spin frequencies
respectively. The tick-marks indicate the \beat\ and \w\,+\,\W\ sideband
frequencies.}
\label{fig:ft}
\end{center}
\end{figure}

Power spectra of these lightcurves are given in
Fig.~\ref{fig:ft}.  The dominant pulses are the 4.85-hr and 20.9-min
orbital and spin pulses (we label the frequencies \W\ and \w\
respectively), and there is also power at the \beat\ and $\w+\W$
sidebands.

Power at the beat period (\beat) is indicative of spin--orbit
interaction in the accretion process. X-ray beat periods are
generally interpreted as resulting from accreting material which
couples directly to the white dwarf magnetosphere, without passing
through an accretion disc (e.g. stream-fed accretion, or
disc-overflow accretion; Hellier \etal1989a; Hellier 1991; Wynn \&\
King 1992).

However, we see approximately equal power at the \beat\ and $\w+\W$
sideband frequencies, which may result simply from an amplitude
modulation of the spin pulse at the orbital timescale (see Warner
1986), rather than from a stream--magnetosphere interaction. To test
this we used the light-curve reconstruction program of H93 to see
whether we could reproduce the light curve using a spin pulse
multiplied by an orbital modulation, without any intrinsic modulation
at the \beat\ period.

The resulting model lightcurve is given in the bottom panel of
Fig.~\ref{fig:curve}, and the power spectrum of the residuals to this
fitted model is shown in the lower panel Fig.~\ref{fig:ft}. The
absence of peaks significantly above the flickering implies that the
model is adequate in reproducing the periodic components of the
lightcurve.  Thus we conclude that there was little or no
disc-overflow accretion occurring at the time of the \emph{XMM-Newton\/}
observations.

Note that Hellier (1991) and B98 reported that the relative amplitude
of the beat and spin pulses change with time, and suggested that the
accretion mode of FO~Aqr changes from periods with significant
disc-overflow accretion to periods where accretion occurs exclusively
via the disc. This phenomenon has also been seen in TX Col (Wheatley
1999).

We also note from Fig.~\ref{fig:ft} that our data show very little
power at the 2\w\ (marked) and 4\w\ (not shown) frequencies, in
contrast to previous data (B98); this indicates substantial changes in
the pulse profile over time.

\section{Spectroscopy}
\label{sec:spectra}

\begin{figure*}
\begin{center}
\psfig{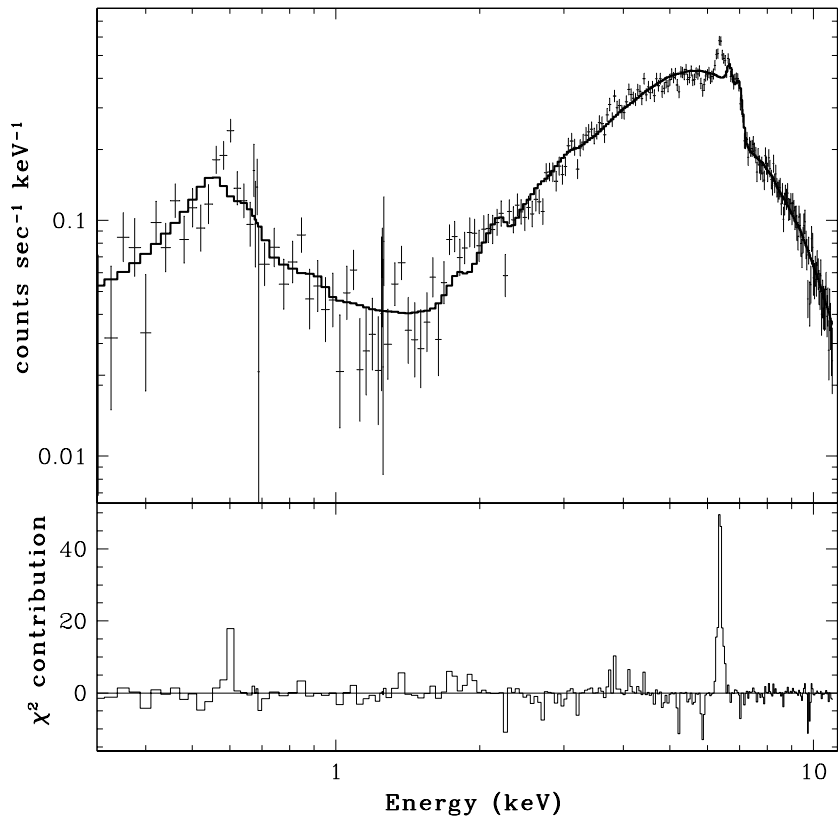}
\caption{Data, model and residuals of the phase-averaged spectrum
from the EPIC-PN camera. The model contains the emitters and
absorbers described in the text, but without the Gaussian components.
There are two large spikes in the $\chi^2$ panel corresponding to the
6.4-keV iron and 0.57-keV oxygen {\sc vii} lines, demonstrating the
need for these components.}
\label{fig:lines}
\end{center}
\end{figure*}

We extracted phase-averaged source and background spectra from the
MOS-1, MOS-2 and PN cameras, and generated response and ancillary
response files using the {\sc sas rmfgen} and {\sc arfgen} tasks.  We
fitted the data from all three cameras simultaneously in {\sc xspec},
making no allowance for calibration systematics between the
instruments. 

The X-ray emission in an IP arises from plasma which is heated to
X-ray temperatures at a stand-off accretion shock, and cools as it
approaches the white dwarf surface (e.g.\ Aizu 1973; Cropper
\etal1999). This can be represented by several \mbox{\sc mekal}
emitters at different temperatures or by using a \mbox{\sc cemekl}
model, which assumes a continuous power-law distribution of \mbox{\sc
mekal} emitters.

A single-temperature {\sc mekal} model (plus absorption) gave a fit
with \rchisq\ = 1.64 and a temperature of 9 keV; adding a second
\mbox{\sc mekal} improved \chisq\ by 18 with temperatures 0.2 keV and
9.3 keV; a third \mbox{\sc mekal} improved the \chisq\ value by a
further 10 (\rchisq\ = 1.63), and gave temperatures of 0.2, 3.0 and 14
keV.  A fourth \mbox{\sc mekal} gave a negligible improvement in
\chisq.

Alternatively, using the {\sc cemekl} model to give a continuous
distribution of temperatures resulted in a comparable \rchisq\ of
1.64.  We use the 3-{\sc mekal} model for phase-resolved analysis
below; note, however, that we do not interpret the highest
temperature as a shock temperature, since this would require the
fitting of a full accretion-column model (e.g.\ Cropper \etal1999)
and is also affected by the likely presence of a reflection component
(e.g.\ Beardmore, Osborne \&\ Hellier 2000). Instead, the 3-{\sc
mekal} model gives an easily-used and flexible representation of the
multi-temperature spectrum for studying spectral changes over spin
phase.

Blackbody emission may be anticipated in an IP, owing to the heating
of the white dwarf surface by the post-shock material, however a
blackbody component in the model made negligible difference to the
\chisq\ value, so was not included.

The spectra show lines of 6.4-keV iron and 0.57-keV oxygen {\sc vii}
that are not reproduced by the 3-{\sc mekal} model (see
Fig.~\ref{fig:lines}). The RGS spectrum confirms that the 0.57-keV
line is a real feature. In what follows we added narrow Gaussian
components to fit these lines.

As is usual in IPs, simple photoelectric absorption was not adequate
to fit the spectrum (giving \rchisq\,=\,3.44 for a simple
photoelectric absorber acting on the 3-{\sc mekal} emitter described
above). Adding a partial covering absorber gave \rchisq\,=\,1.87\ and
a second such component reduced the \rchisq\ to 1.63.  The fitted
parameters of the 3-{\sc mekal}, 3-absorber model are given in
Table~\ref{tab:comps}. Note that the amount of absorption changes over
spin phase (Section~\ref{sec:spin}) so the parameters fitted to the
phase-averaged spectrum will be a weighted average over spin phase.

The above \chisq\ values make no allowance for normalisation or
calibration systematics between the three instruments. As a test, we
have also allowed the parameters of the the 3-{\sc mekal}, 3-absorber
model to optimise for each instrument separately. This reduces the
overall \rchisq\ from 1.63 to 1.03.  When we report the phase-resolved
analysis we quote $\Delta$\chisq\ values between competing models.
Although the absolute \chisq\ is lower when allowing
the parameters to optimise for each instrument, the $\Delta$\chisq\
values are similar. 

One result of the above modelling is the finding of a best-fitting
abundance of only 0.12 solar. This results from a relative lack of
lines in the data compared to that predicted by the {\sc mekal} code,
which assumes that the plasma is optically thin and collisionally
ionized. Thus, the different line emission could equally be due to
optical depth effects or photo-ionization (see, e.g., Mukai \etal\
2003, for such effects in IP spectra). To investigate whether the
abundance is genuinely low, or whether the line emission is
suppressed, we added a bremsstrahlung emitter for each of the
\mbox{\sc mekal} emitters with their temperatures fixed to those of
the \mbox{\sc mekal}s; and fixed the abundance to 1 (i.e. solar).
This allowed a fit to be obtained which was as good as that with a low
abundance, but the difference in \chisq\ was too low (\til5) to
indicate which of these models is most physical. Applying this model
to the higher resolution RGS data showed that either method reproduced
the data, provided the abundance was above \til0.15. We use the
variable-abundance model in what follows.

\begin{table}
\begin{center}
\begin{tabular}{llcl}
\hline
Component &  Parameter       & Value                 & Error \\ 
          &  (Units) \\
\hline
Absorption      &  nH (cm$^{-2}$)  & $1.02 \tim{21}$ & $^{+0.81\tim{21}}_{-0.05\tim{21}}$\\
Partial Absn.&  nH (cm$^{-2}$)  & $1.90 \tim{23}$ & $^{+0.12\tim{23}}_{-0.09\tim{23}}$\\
          &  CvrFract        & 0.855           & $^{+0.021}_{-0.026\tim{-2}}$\\
Partial Absn.&  nH (cm$^{-2}$)  & $5.71 \tim{22}$ & $^{+0.45\tim{22}}_{-0.41\tim{22}}$\\
          &  CvrFract        & 0.980           & $^{+0.004}_{-0.005}$ \\
Gaussian  &  Energy (keV)          & 6.40            & {\sc frozen}\\
          &  Normalisation         & $9.23 \tim{-5}$ & $^{+0.695\tim{-5}}_{-1.41\tim{-5}}$\\
          &  Eq. width (eV)     & 129             &\\
Gaussian  &  Energy (keV)          & 0.587           & $^{+0.031}_{-0.029}$\\
          &  Normalisation         & $6.84 \tim{-3}$ & $^{+4.69\tim{-3}}_{-4.54\tim{-3}}$\\
          &  Eq. width (eV)     & 22.9            &\\
Mekal     &  kT (keV)              & 0.176           & $^{+0.024}_{-0.026}$\\
          &  Abundance             & 0.121           & $^{+0.028}_{-0.023}$ \\
          &  Normalisation         & 0.160           & $^{+0.365}_{-0.094}$\\
Mekal     &  kT (keV)              & 3.00            & $^{+1.48}_{-0.73}$ \\
          &  Normalisation         & $3.32 \tim{-2}$ & $^{+1.87\tim{-2}}_{-1.51\tim{-2}}$ \\
Mekal     &  kT (keV)              & 13.9            & $^{+6.5}_{-3.1}$\\
          &  Normalisation         & $4.02 \tim{-2}$ & $^{+0.88\tim{-2}}_{-1.06\tim{-2}}$\\
\hline
\end{tabular}
\caption{Model components and parameters fitted to the phase-averaged
spectrum. The errors show the 90\% confidence limit, according to
formal statistics, making no allowance for calibration uncertainties
between the three detectors.}
\label{tab:comps}
\end{center}
\end{table}

\section{Orbital modulation}
\label{sec:orb}

X-ray observations of FO~Aqr often show a prominent modulation on the
orbital cycle (e.g.\ N92), as is obvious in our UV and X-ray data
(Fig.~1).  The standard interpretation is that FO~Aqr is a
high-inclination system in which disc material, particularly at the
stream--disc impact region, periodically obscures the line of sight
to the white dwarf (e.g.\ H93).  Note that Hellier, Mason \&\ Cropper
(1989b) reported a grazing optical eclipse in FO~Aqr, implying an
inclination of \til65\deg$\!\!$, although there is no sign of the
white dwarf itself being eclipsed.

Other possible causes of an orbital modulation include stream-fed
accretion in which the accretion footprints are linked to orbital
phase (N92). This, though, is less able to explain our dataset given
the absence of the X-ray beat modulation that would also result
(Section~\ref{sec:fts}).

\begin{figure}
\begin{center}
\psfig{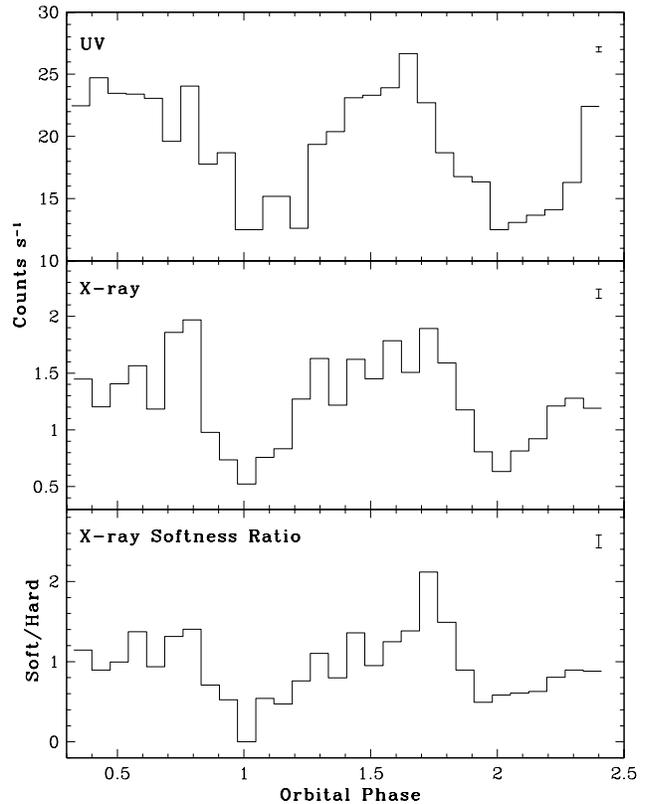}
\caption{Lightcurves binned at 1254 s, thus removing the spin
modulation. The data are not folded but are plotted against phase.
The top two panels show the UV (OM) and the X-ray (MOS 1+2) data. The
bottom panel shows the MOS 1+2 softness ratio, defined as
\mbox{0.2--4\,/\,6--12} keV.  Typical error bars are shown. Phase
zero occurs at HJD 2452041.806, and has been defined as the minimum
of the UV pulse, estimated by eye.}
\label{fig:orb}
\end{center}
\end{figure}

Fig.~\ref{fig:orb} shows a softness-ratio plot of the orbital modulation,
constructed using 1254-s bins to smooth out the spin pulse (we show
the data unfolded since we have only \til2.1 orbital cycles of data).
The softness ratio closely follows the lightcurve, implying that the
spectrum gets harder in the minima, supporting the absorption idea.

To investigate further we extracted spectra from the three X-ray 
cameras in the orbital dip (\p\,=\,0.85--1.2) and out of the dip
(\p\,=\,1.2--0.85).  Fitting the spectra with the model discussed
in Section~\ref{sec:spectra}, we found that the spectra were
adequately reproduced by a model in which only the absorption changed
during the dip.  This required increased partial-covering absorbers
with a density near 3\tim{23} cm$^{-2}$, and gave a best-fitting
\rchisq\ of 1.37 (with much of the excess \chisq\ arising from
systematics between the three X-ray cameras).  Allowing other
parameters to vary in addition to the absorption improved the \chisq\
by only 18 (for $\nu = 2164$), which is not significant.

From the softness ratio and spectral modelling, and the absence of
beat-cycle modulations, we conclude that the orbital modulation is a
`dipping' event resulting from absorption by stream or disc material.
The material must extend to \sqiggt 25\deg\ above the plane, given
the inclination quoted above, and must extend for at least 0.4 in
orbital phase. This suggests that the stream/disc interaction region
is quite extensive.

\section{Spin Modulation}
\label{sec:spin}

\begin{figure}
\begin{center}
\psfig{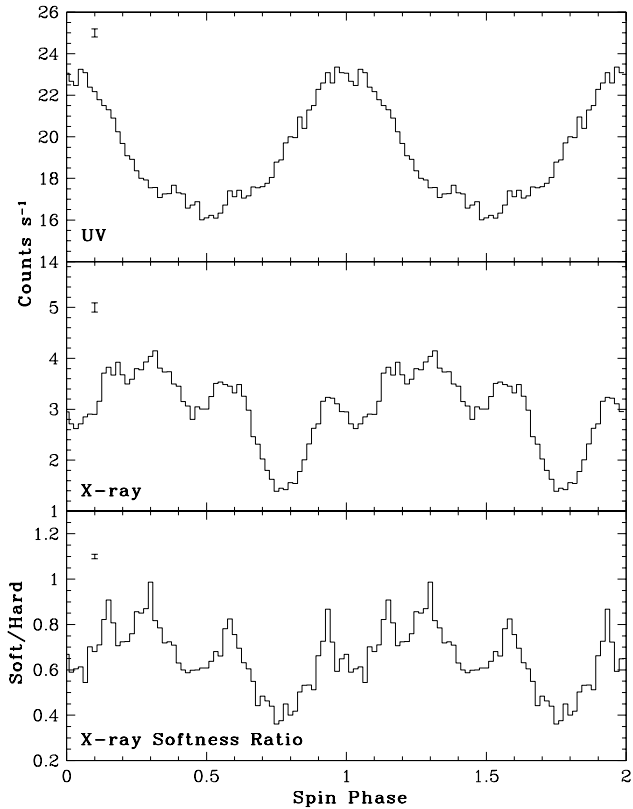}
\caption{Spin pulse profiles, folded on \w=1254.4459 s. Phase zero
occurs at HJD 2452041.858. The upper panel shows the UV (OM) data.
The central panel shows the X-ray (MOS+PN) data, and the
\mbox{0.2--4\,/\,6--12} keV softness ratio is in the lower panel.
Typical errors are given.}
\label{fig:spin}
\end{center}
\end{figure}

Fig.~\ref{fig:spin} shows the UV and summed X-ray lightcurves folded
on the 1254-s spin period, along with the softness ratio of the X-ray
data. We have phased our data such that spin-phase zero occurs at the
maximum of the UV pulse, located by fitting a sine curve to those
data. This phase zero corresponds to 0.23 on the Williams (2003)
ephemeris.

The X-ray pulse profile shows a main dip (at phase 0.75 on our plots)
followed by a `notch' feature (phases 0.0--0.1), which concurs with
previous reports by N92, H93, and B98. The softness ratio of the
X-ray spin pulse follows the morphology of the total flux level.

\subsection{Spectral analysis}
\label{sec:specspin}

To investigate the spin-pulse features noted above we extracted
spectra from all three X-ray cameras from spin `maximum' (\p\,\til
0.12--0.36), `minimum' (\p\,\til 0.67--0.87), and the `notch'
(\p\,\til 0.99--1.11). We applied the three-\mbox{\sc mekal} model
described in Section~\ref{sec:spectra} simultaneously to `maximum'
and `minimum', allowing the parameters to vary between regions, and
then fixing the emission or absorption, in order to investigate the
probable cause of the changes. We then repeated this process between
the `maximum' and `notch' phase regions. {\sc xspec} could not always
constrain the \mbox{\sc mekal} temperatures, so we froze these at
0.1, 2 and 30 keV as this had negligible effect on the fit quality.
We also tied the abundance across phase regions, since it had a
minimal effect on the fit quality ($\Delta\chisq\til5$), and
fixed the 0.57-keV line at its rest energy with zero width.

Fitting the maximum and minimum phase regions simultaneously without
requiring any model parameters to be the same (with the exceptions
noted above) yielded a best-fitting model of \rchisq\ = 1.2 ($\nu =
1161$). Note, again, that much of the \chisq\ comes from calibration
uncertainties; optimising for each instrument separately gave \rchisq
= 0.87. Since these systematics dominate \chisq, true uncertainties are
much greater than the probabilities drawn from
formal statistics; thus we present our results using 
$\Delta\chisq$ values rather than probabilities.

Constraining the model so that the absorption could not vary between
maximum and minimum made the fit worse by $\Delta\chisq=53$; if
instead the absorption is allowed to vary and the emission parameters
are fixed, the fit is worse by $\Delta\chisq=29$.  This suggests that
the minimum is mostly the result of absorption.

We then fitted the maximum and notch simultaneously: allowing only
the absorption to vary worsened the fit by $\Delta\chisq=39$ compared
to an unconstrained model, whereas allowing only the emission
parameters to vary worsened the fit by only $\Delta\chisq=4$.  From
this we conclude that the notch is probably not an absorption event,
but is caused by a change in the underlying emission, such as the
occultation of parts of the accretion region.

\section{Discussion}
\label{sec:discspin}

FO~Aqr has a complex X-ray spin-pulse profile. The main feature is a
dip, centered at \p\,\til 0.75, which is caused by increased
absorption (section~\ref{sec:specspin}). Strong absorption dips are
characteristic of IPs, and in the standard model occur when the
accretion curtains of infalling material obscure the line of sight to
the accretion regions (e.g.\ Rosen, Mason \&\ C\'ordova 1988;
Hellier, Cropper \&\ Mason 1991; Kim \&\ Beuermann 1995).

Our data also show a smaller flux reduction, the `notch' at \p\,\til
0.0. This notch is persistent in FO~Aqr's X-ray lightcurve (e.g.\
N92, B98) and is often more prominent than in our data.  Our spectral
analysis suggests that this is the disappearance of part of the
emitting region, rather than an absorption dip.

N92 attributed the notch to the disappearance over the white-dwarf
limb of a `hot-spot' within the accretion region at the upper pole.
H93 suggested that it is probably the upper pole itself that
disappears: the width of the notch implies a region covering 0.001 of
the white-dwarf area (N92), which is in line with the estimate of
$<$\,0.002 for the size of the accreting poles in the eclipsing
system XY~Ari (Hellier 1997a).

Our simultaneous far-UV lightcurve shows a nearly-sinusoidal spin
pulse with a maximum coincident with the notch (\p\,\til 0.0).  In the
standard accretion-curtain model, the optical and UV maximum occurs
when the upper pole is on the far side of the white dwarf, furthest
from the observer, and the accretion curtains are best displayed
(Hellier \etal1987). This is consistent with our interpretation of the
notch.  We do, though, require an asymmetry in the visibility of the
upper and lower poles, otherwise the appearance of the lower pole
would compensate for the disappearance of the upper pole and fill in
the notch.  This point was previously noted by B98, who showed that
the notch lightcurve could be reproduced if the magnetic dipole were
offset by 0.15 white-dwarf radii from the centre.

Given the above, a simple model would predict that the absorption dip
caused by the upper accretion curtain passing through the line of
sight would occur half a cycle later, at \p\,\til 0.5.  In fact, it
occurs at \p\,\til 0.77, late by 0.27 cycles.  A simple shift in
phase, such as accreting only along field lines trailing the pole,
would not by itself explain the quarter-cycle discrepancy between
phases deduced from the UV and the X-ray. Thus we conclude that the
field lines are not an undistorted dipole, but are twisted, being
swept backwards by a quarter of a cycle.

We can use B98's compilation of FO~Aqr lightcurves to investigate
whether the amount of distortion is variable over time.  The notch is
prominent in \emph{Ginga\/} data taken in 1988 and 1990, and in both
cases implies an accretion-curtain lag comparable to the amount we
see in our data.  The notch is either less obvious or absent in
\emph{EXOSAT\/} data from 1983 and 1985, and in \emph{ASCA\/} data
from 1993, but the statistical quality of those lightcurves is much
lower. Thus, where both a notch and a dip are seen, the
phase-difference between them is compatible with a constant lag, and
there is no evidence for episodes when the accretion-curtain twist is
markedly different.

\subsection{Comparison with PQ~Gem} 

While many intermediate polars have quasi-sinusoidal spin pulses,
PQ~Gem, like FO~Aqr, has a more complex pulse (e.g.\ Mason 1997).
However, in that star it is proposed that the accreting field lines
\emph{precede\/} the magnetic pole, the opposite to our conclusion for
FO~Aqr, based on analysis of the X-ray pulse (Mason 1997), optical
polarimetry (Potter \etal1997), and optical spectroscopy (Hellier
1997b).

A possible line of enquiry is whether the twists in the accretion
curtains -- opposite in the two stars -- are related to the accretion
torques on the white dwarfs.  In PQ~Gem the white dwarf is spinning
down on a $2.4\times10^{4}$-yr timescale (Mason 1997).  FO~Aqr's
white dwarf has shown a more complex history, having been spinning
down between 1981 and 1988, after which it changed to spinning up
(Williams 2003).

The spin-pulse profiles of FO~Aqr have varied over that time period
(e.g.\ B98), however there is no simple change in the relative
locations of dip and notch that could be attributed straightforwardly
to the change from spin down to spin up.  Thus, although it is
tempting to note that in the spinning-down PQ~Gem the field lines
precede the magnetic pole, whereas in the (currently) spinning-up
FO~Aqr they trail the magnetic pole, any deductions from this are not
straightforward.

Ghosh \&\ Lamb (1979) argue that the net accretion torque on the
white dwarf combines the spin-up torque from the angular momentum of
the accreted material with the torques of field lines dragging
through the accretion disc.  The expected result is a spin period
which oscillates around an equilibrium period, undergoing phases of
spin up and spin down.  The field--disc torques could be the cause of
the twists in the accretion curtains.

Mason (1997) and Potter \etal(1997), on the basis of optical
polarimetry, proposed that in PQ~Gem the accreting material threads
field lines outside the corotation radius (where the field lines are
moving faster than the disc material) and that this leads to the
spin-down seen in that star.  However, Mason (1997) notes that
accretion from outside the corotation radius contradicts the standard
theory of how an accretion disc interacts with a magnetic field
(e.g.\ Ghosh \&\ Lamb 1979).  Our results from FO~Aqr mean that the
theory must also be adaptable to explain accretion along trailing
field lines, and also the puzzling fact that the lines appear to
trail both in episodes of spin-up and of spin-down.  Thus any
interpretations of the field twists seen in PQ~Gem and FO~Aqr are
currently uncertain.

\section{Conclusions}
\label{sec:conc}
\newcounter{conc}

Analysis of \emph{XMM-Newton\/} data on FO~Aqr has shown that at the
time of our observations: 

\begin{list}{(\arabic{conc})}{\usecounter{conc}\setlength{\rightmargin}{\leftmargin}}
\item{There was negligible disc-overflow accretion.}

\item{An orbital dip was probably caused by absorption of the X-ray
emission by vertical structure in the accretion disc.}

\item{A narrow `notch' in the X-ray spin pulse is likely to be caused
by the upper accretion region disappearing over the limb of the white
dwarf.}

\item{The UV spin pulse is quasi-sinusoidal with a maximum 
coincident with the notch. This will occur when the base of the upper
pole is furthest from us, and accords with the
accretion curtain model for UV/optical spin pulses.}

\item{A broad dip in the X-ray pulse is caused by absorption, most
likely occurring when the outer regions of the upper accretion
curtain  cross our line of sight to the white dwarf.}

\item{The phase difference between the broad dip and the notch is
0.23, whereas a simple accretion-curtain model would predict 0.5.
This implies that the curtains are twisted, with the outer parts
trailing the magnetic pole.}

\item{The finding of a trailing curtain contrasts with reports
that the accretion curtains lead the magnetic poles in PQ~Gem.}

\item{Although PQ~Gem's white dwarf is spinning down, whereas that in
FO~Aqr is (currently) spinning up, there appears to be  no simple
relation between the period change and whether the  curtain trails or
leads.}

\end{list}

\label{lastpage}
\end{document}